\let\accentvec\vec
\documentclass{llncs}

\let\vec\accentvec
\usepackage{tikz}
\usepackage{amsmath}
\usepackage{amssymb}
\usepackage{fancybox}
\newcommand{\COMMENTED}[1]{{}}
\newcommand{\junk}[1]{\COMMENTED{#1}}
\usetikzlibrary{patterns}
\begin{document}
\title{A combinatorial algorithm for all-pairs shortest paths
in directed vertex-weighted graphs with applications
to disc graphs}
\author{
Andrzej Lingas \inst{1}
\and
Dzmitry Sledneu \inst{2} 
\institute{
Department of Computer Science, Lund University, 22100
Lund. \texttt{Andrzej.Lingas@cs.lth.se. Fax +46 46 13 10 21}
\and
The Centre for Mathematical Sciences, Lund University, 22100 Lund, Sweden.
\texttt{Dzmitry.Sledneu@math.lu.se}
}}
\date{}
\maketitle
\begin{abstract}
We consider the problem of computing
all-pairs shortest paths in a directed graph
with real weights assigned to vertices. 

For an $n\times n$ $0-1$ matrix $C,$ let $K_{C}$ be the complete
weighted graph on the rows of $C$ where the weight of an edge between
two rows is equal to their Hamming distance.  Let $MWT(C)$ be the weight
of a minimum weight spanning tree of $K_{C}.$ 

We show that the
all-pairs shortest path problem for a directed graph $G$ on $n$
vertices with nonnegative
real weights and adjacency matrix $A_G$ can be solved by a
combinatorial randomized algorithm in time\footnote{
The notation $\widetilde{O}(\ )$ suppresses 
polylogarithmic factors and 
$B^t$ stands for the transposed matrix $B.$} 
$$\widetilde{O}(n^{2}\sqrt {n +
\min\{MWT(A_G), MWT(A_G^t)\}})$$

As a corollary, we conclude that the transitive
closure of a directed graph $G$ can be computed
by a combinatorial randomized algorithm
in the aforementioned time.
\junk{$\widetilde{O}(n^{2}\sqrt {n +
\min\{MWT(A_G), MWT(A_G^t)\}})$ }

We also conclude that the
all-pairs shortest path problem for 
uniform disk
graphs, with nonnegative real vertex weights,
induced by point sets of
bounded density within a unit
square can be solved in time 
$\widetilde{O}(n^{2.75})$. 
\end{abstract}

\section{Introduction}

The problems of finding shortest paths and determining
their lengths are fundamental in algorithms. They
have been extensively studied in algorithmic graph
theory. A central open question in this area is if
there is a substantially subcubic in the number
of vertices algorithm for the all-pairs shortest
path problem for directed graphs with real edge 
weights (APSP) in the addition-comparison
model \cite{VW10,Z01}. For several special cases of
weights and/or graphs substantially subcubic
algorithms for the APSP problem are known \cite{AGM,C07,S,Y,Z02,Z01}.
However, in the general case the fastest
known algorithm due to Chan \cite{C07} (see also \cite{C08})
runs in time $O(n^3\log^3\log n/\log^2n)$,
achieving solely a moderate polylogarithmic
improvement over the $O(n^3)$ bound yielded
by Floyd-Warshall and Johnson's algorithms \cite{AHU,Z01}.
\junk{(Marginally, the APSP problem in unweighted random
directed graphs or even in complete directed graphs
with random edge weights can be solved in expected time 
$\widetilde {O} (n^2)$ \cite{PSSZ10,SS1}.)}

The situation is different for directed graphs with
real vertex weights. Recently, Chan has shown
that the APSP problem for the aforementioned
graphs can be solved in time $O(n^{2.844})$ \cite{C07}
and Yuster has slightly improved the latter
bound to $O(n^{2.842})$ by using an improved bound
on rectangular multiplication \cite{Y}.

The basic tool in achieving substantially subcubic
upper bounds on the running time for the APSP
for directed graphs with constrained edge weights
or real vertex weights are the fast algorithms
for arithmetic square and rectangular
matrix multiplication \cite{CW,HP98}. One typically
exploits here the close relationship between
the APSP problem and the so called distance or
$(\min , +)$ product \cite{AGM,S,VW10,Y,Z01,Z02}. 

Unfortunately, these fast algorithms for matrix multiplication,
yielding equally fast algorithms for Boolean matrix product, are based
on recursive algebraic approaches over a ring
difficult to implement. Thus,
another central question in this area is whether or not there is a
substantially subcubic {\em combinatorial} 
(i.e., not relaying on ring algebra) algorithm for the Boolean product of
two $n\times n$ Boolean matrices \cite{BW,R85,VW10}. 
Again, the fastest
known combinatorial algorithm for Boolean matrix product due to Bansal
and Williams \cite{BW} running in time $O(n^3\log^2 \log n/\log^{9/4}
n)$ achieves solely a moderate polylogarithmic improvement over the
trivial $O(n^3)$ bound.  On the other hand, several special cases of
Boolean matrix product admit substantially subcubic combinatorial algorithms
\cite{BL,GL,L}.

In particular, Bj\"orklund et al. \cite{BL} provided a combinatorial
randomized algorithm for Boolean matrix product which is substantially subcubic
in case the rows of the first $n\times n$ matrix or the columns of the
second one are highly clustered, i.e., their minimum spanning tree in
the Hamming metric has low cost.  More exactly, their algorithm runs
in time $\tilde{O}(n(n + c)),$ where $c$ is the minimum of the costs of
the minimum spanning trees for the rows and the columns, respectively,
in the Hamming metric.  It relies on the fast 
Monte Carlo methods for computing an
approximate minimum spanning tree in the $L_1$ and $L_2$ metrics given
in \cite{IM,IST}.

The assumption that the input directed graph is highly clustered in
the sense that the minimum spanning tree of the rows or columns of its
adjacency matrix in the Hamming metric has a subquadratic cost does not
yield any direct applications of the algorithm of Bj\"orklund et
al. \cite{BL} to shortest path problems, not even to the transitive
closure.  The reason is that the cost of the analogous minimum
spanning tree can grow dramatically in the power graphs
\footnote{In the $i$-th power graph there is an edge from $v$ to
$u$ if there is a path composed of at most $i$ edges from $v$ to $u$
in the input graph.}
of the input graph.
In particular, we cannot obtain directly
an upper time-bound
on the transitive closure of Boolean matrix
corresponding to that
for the Boolean matrix product from \cite{BL}
by applying the asymptotic equality between the
time complexity of matrix product over a 
closed semi-ring and that of its
transitive closure over the semi-ring due to Munro \cite{M71}.
The reason is the dependence of the upper bound from \cite{BL} 
on the cost of the minimum spanning tree.

In this paper, we extend the idea of the method from \cite{BL} 
to include a mixed product of a real matrix with a Boolean one.
We combine the aforementioned extension with
the ideas used in the design of subcubic algorithms for important
variants of the APSP problem \cite{AGM,Z02}, in particular those for
directed graphs with vertex weights \cite{C07,Y}, to obtain not only a
substantially subcubic combinatorial algorithm for the transitive
closure but also for the APSP problem in highly clustered directed
graphs with real vertex weights.

For an $n\times n$ $0-1$ matrix $C,$ let $K_{C}$ be the complete
weighted graph on the rows of $C$ where the weight of an edge between
two rows is equal to their Hamming distance.  Let $MWT(C)$ be the weight
of a minimum weight spanning tree of $K_{C}.$ We show that the
all-pairs shortest path problem for a directed graph $G$ on $n$
vertices with nonnegative
real weights and an adjacency matrix $A_G$ can be solved by a
combinatorial randomized algorithm in  $\widetilde{O}(n^{2}\sqrt {n +
\min\{MWT(A_G), MWT(A_G^t)\}})$ time.
It follows in particular that the transitive closure
of a directed graph $G$ can be computed by a
combinatorial randomized algorithm in the aforementioned time.
 
Our algorithms are of Monte Carlo type and by increasing the
polylogarithmic factor at the time bounds,
the probability that they return
a correct output within the bounds can
be amplified to $1-\frac1{n^{\alpha}},$ where $\alpha \ge 1.$

{\em Since there are no practical or combinatorial substantially subcubic-time
algorithms not only for the APSP problem
but even for the transitive closure problem for arbitrary
directed graphs at present,
our simple  adaptive method might be a potentially interesting alternative
for a number of graph classes.}  

As an example of an application of our method, we consider
the APSP problem for uniform disk graphs, with nonnegative
real vertex weights,
induced by point sets of bounded
density within a unit square. We obtain a
combinatorial algorithm for this problem running in
time $O(\sqrt rn^{2.75})$, where $r$ is the radius
of the disks around the vertices 
in a unit square. 

The recent interest in disk graphs,
in particular uniform disk graphs, stems from
their applications in wireless networks. In this context,
the restriction to point sets of bounded density
is quite natural. In \cite{FK06}, F\"urer and Kasiviswanathan
provided a roughly $O(n^{2.5})$-time preprocessing
for {\em approximate} $O(\sqrt n)$-time distance queries
in arbitrary disk graphs.

\junk{Note here that the aforementioned
bounds for unweighted random directed graphs \cite{SS1}
or for complete directed graphs with random weights \cite{PSSZ10}, 
respectively, do not apply in this case.}

Our paper is structured as follows.
In the next section, we show a reduction of the APSP
problem for directed graphs with real vertex-weights
to a mixed matrix product of a distance matrix over
reals with the $0-1$ adjacency matrix. In Section 3, we present
an algorithm for such a mixed product which generalizes
that for the Boolean matrix product from \cite{BL} and
runs in subcubic time if the input $0-1$ matrix is highly
clustered. By combining the results of Sections 2,3, we
can derive our main results in Section 4. 
In the next section, we present the application of
our method to uniform disk graphs induced by point sets
of bounded density. We conclude with
final remarks.

\section{A Reduction of APSP to Mixed Matrix Products}

\subsection{The APSP problem}

Formally, the All-Pairs Shortest Paths problem (APSP)
in a directed graph 
$G=(V,E)$ with real weights $w(v)$
associated to vertices $v\in V$
is to compute the $|V|\times |V|$ distance matrix $D_G$
such that $D_G(v,u)$ is the distance $\delta_G(v,u)$
from $v$ to $u$ in $G,$ i.e.,
the minimum total weight
of vertices on a path from $v$ to $u$ in $G.$
An additional goal of the APSP problem is to
compute a concise data structure representing
the shortest paths.

Note that $\delta_G(v,u)$
is equal to the minimum total weight
of inner vertices on a path from $v$ to $u$ in $G$
increased by the weights of $v$ and $u.$

We shall assume $|V|=n$ throughout the paper.

For $i=0,1,...,n-1,$
let $\delta_G^i(v,u)$ be the distance
from $v$ to $u$ on paths consisting of at
most $i$ edges, i.e.,  
the minimum total weight
of vertices on a path 
from $v$ to $u$ having at most $i$
edges in $G.$ Next, let $D_G^i$ be the
 $|V|\times |V|$ matrix such that
$D_G^i[v,u] $ is equal to $\delta_G^i(v,u)$.

For convention,
we assume $\delta_G^0(v,v)=0$ and $\delta_G^0(v,u)=
+\infty$ for $v\neq u.$
Hence, $D_G^0$ has zeros on the diagonal
and $+\infty$ otherwise. In $D_G^1,$ all the
entries $D_G^1[v,u]$ where $(v,u)\in E$
are set to $w(v)+w(u)$ instead of $+\infty$.
Thus, both $D_G^0$ and $D_G^1$ can be easily
computed in time $O(n^2).$

\subsection{Mixed Matrix Products}

Let $A$ be an $n\times n$ matrix over $R\cup 
\{ +\infty \},$ and let $B$ be an $n\times n$ matrix
with entries in $\{0,1\}.$
The {\em mixed right product} $C$ of $A$
and $B$ is defined by

$$C[i,j]=\min \{ A[i,k]|1\le k\le n\ \& \ B[k,j]=1\}\cup \{ +\infty \}$$.

If $C[i,j]\neq +\infty$ then the index $k$ 
such that $C[i,j]=A[i,k]$ (and thus $B[k,j]=1$)
is called a witness for $C[i,j].$
Analogously, the 
{\em mixed left product} $C'$ of $B$ and  $A$ 
is defined by 
$$C'[i,j]=\min \{ A[k,j]| 1\le k\le n\ \& \ B[i,k]=1\}\cup \{ +\infty \},$$
and if $C'[i,j]\neq +\infty$ then the index $k$
such that $C'[i,j]=A[k,j]$ is called 
a witness for $C'[i,j].$
\junk
{For two $n\times n$ matrices $A$ and $B$
over $R\cup \{\infty\}$,
the {\em distance product} $C=A*B$ is defined
by $C[i,j]=\min_{k=1}^n A[i,k]+B[k,j].$}

An $n\times n$ matrix $W$ such that 
whenever $C[i,j]\neq +\infty$ 
then $W[i,j]$ is a
witnesses for $C[i,j]$
is called a witness matrix for the 
right mixed product of $A$ and $B.$
Analogously, we define a witness matrix for the left mixed product
of $B$ and $A.$

\subsection{The Reduction}

Let $A_G$ denote the $n\times n$ adjacency matrix of
$G=(V,E),$ i.e., $A_G[v,u]=1$ iff $(v,u)\in E.$

\begin{lemma}\label{lem:i+1}
For an arbitrary $i\in \{ 0,1,...,n-2\},$ $D_G^{i+1}$ can be
computed 
on the basis of $D_G^i$ and
the right mixed product of $D_G^i$ with $A_G$
or $D_G^i$ and the left mixed  product of $A_G$ with $D_G^i$
in time $O(n^2).$
\end{lemma}

\begin{proof}
It is sufficient to observe that for any pair $v,\ u$ of vertices
in $G$, $D_G^{i+1}[v,u]$ is equal to 
$$\min \{ D_G^i[v,u],\min \{D_G^i[v,x]+w(u)|
1\le x\le n\ \& \ A_G[x,u]=1\} \cup \{ +\infty \}\}$$
Symmetrically, $D_G^{i+1}[v,u]$ is equal to
$$\min \{ D_G^i[v,u],\min \{D_G^i[x,u]+w(v)|
1\le x\le n\ \& \ A_G[v,x]=1\} \cup \{ +\infty \}\}$$\qed
\end{proof}

The following lemma follows the general strategy
used to prove Theorem 3.4 in \cite{C07}.

\begin{lemma}\label{lem: mix}
Let $G$ be a directed graph $G$ on $n$ vertices
with nonnegative real vertex weights.
Suppose that the right (or left)
mixed product of an $n\times n$ matrix over $R\cup \{ +\infty\}$
with the adjacency matrix $A_G$ of $G$
along with the witness matrix
can be computed in time $T_{mix}(n)=\Omega (n^2).$
The APSP problem for $G$ 
can be solved in time $\widetilde {O} (n^{1.5}\sqrt {T_{mix}(n)}).$
\end{lemma}
\begin{proof}
We begin by computing $D_G^{t-1}$ for some $t\in [2,...,n]$
which will be specified later. By Lemma \ref{lem:i+1}
this computation
takes time $O(tT_{mix}(n)).$

It remains to determine distances between pairs
of vertices where any shortest path consists of at least $t$
edges. For this purpose,
we determine a subset $B$ of $V,$ the so called
bridging set \cite{Z02}, hitting all the aforementioned long paths.
We apply the following fact to $l=t$ and
sets of $t$ vertices on 
shortest consisting of exactly $t-1$ edges,
similarly as in \cite{AGM,C07,Y,Z02}.
\par
\vskip 5pt
\noindent
{\bf Fact 1}. {\em Given a collection of $N$ subsets of
$\{1,...,n\},$ where each subset has size exactly $l,$ we can find a
subset $B$  of size $O((n/l)\log n)$ that hits all subsets
in the collection in time $O(Nl).$}
\par
\vskip 5pt
\noindent
Since our application of Fact 1 is analogous to those
in \cite{AGM,C07,Y,Z02}, we solely sketch it
referring the reader for details to the aforementioned
papers.

Note that for each pair  $v,\ u,$  of vertices for which
any shortest path has at least $t$ edges there is a pair
$v',u'$ of vertices on a shortest path
from $v$ to $u$ such that any shortest path from $v'$
to $u'$ has
exactly $t-1$ edges. For all such pairs $v',u',$ we can
find a shortest path on $t-1$ edges, and thus on $t$
vertices, by backtracking on the computation
of $D_G^{t-1}$ and using witnesses for the mixed products.
In total, we generate $O(n^2)$ such paths on $t$ vertices
in time $O(tn^2).$ The application of Fact 1 also takes time
$O(tn^2).$

Next, we run Dijkstra's single-source shortest path algorithm
\cite{AHU}
for all vertices in the bridging set $B$ in the input
graph $G$ and in the graph resulting from reversing
the direction of edges in $G.$ In this way, we determine
$D_G[v,u]$ for all pairs $(v,u)\in (B\times V )\cup (V\times B).$

Now, it is sufficient for all remaining
pairs $(v,u)$ in $V\times V$ to set
$$D_G[v,u]=\min \{ D_G^{t-1}(v,u), \min_{b\in B} \{ D_G[v,b]+D_G[b,u]-w(b)\}$$
in order to determine the whole $D_G.$ 

The computation of $D_G^{t-1}$ takes $O(tT_{mix}(n))$ time
which asymptotically is not less than the $O(tn^2)$ time taken
by the construction of the bridging set.
The runs of Dijkstra's algorithm and 
the final computation of $D_G$
require $\widetilde {O} (\frac nt n^2)$ time. 
By setting $t=\sqrt {\frac {n^3}{T_{mix}(n)}}$, we obtain
the lemma.\qed
\end{proof}

\section{Fast Computation of the Mixed Products for Clustered Data}

Our algorithm for the right (or, left) mixed product relies
on computation of an approximate minimum spanning
tree of the columns (or rows, respectively)
of the Boolean input matrix in the Hamming metric.

\subsection{Approximate Minimum Spanning Tree in High Dimensional Space}

For $c \ge 1$ and a finite set $S$ of points in a metric space,
a {\em $c$-approximate minimum spanning tree for $S$}
is a spanning tree in the complete weighted graph on $S,$
with edge weights equal to the distances between the endpoints,
whose total weight is at most $c$ times the minimum.

In \cite{IM} (section 4.3) and \cite{I} (section 3), 
Indyk and Motwani in particular
considered 
the bichromatic $\epsilon$-approximate
closest pair problem for $n$ points
in $R^d$ with integer coordinates in
$O(1)$ under the $L_p$ metric, $p\in \{ 1,2\}.$
They showed that there is a 
dynamic data structure for
this problem which supports insertions,
deletions and queries in time
$O(dn^{1/(1+\epsilon)})$ and requires
$O(dn +n^{1+1/(1+\epsilon)})$-time preprocessing.
In consequence, by a simulation
of Kruskal's algorithm they deduced
the following fact.
\par
\vskip 5pt
\noindent
{\bf Fact 2}. {\em For $\epsilon >0,$ 
a $1+\epsilon$-approximate minimum spanning
tree for a set of $n$ points
in $R^d$ with integer coordinates in
$O(1)$ under the $L_1$ or $L_2$
metric can be computed by a Monte Carlo
algorithm 
in time $O(dn^{1+1/(1+\epsilon)}).$}
\par
\vskip 5pt
\noindent
In \cite{IST} Indyk, Schmidt and Thorup reported
even slightly more efficient (by a poly-log factor)
reduction of the problem of finding a
$1+\epsilon$-approximate minimum spanning
tree to the bichromatic $\epsilon$-approximate
closest pair problem via an easy simulation
of Prim's algorithm.

Note that the $L_1$ metric for
points in $R^n$ with $0,\ 1$-coordinates
coincides with the $n$-dimensional Hamming metric.
Hence, Fact 2 immediately yields the following
corollary.

\begin{corollary}\label{cor: Ham}
For $\epsilon >0,$ 
a $1+\epsilon$-approximate minimum spanning
tree for a set of $n$ $0-1$ strings of length
$n$ under the Hamming
metric can be computed by a Monte Carlo algorithm 
in time $O(n^{2+1/(1+\epsilon)}).$
\end{corollary}

\subsection{The Algorithm for Mixed Matrix Product}

The idea of our combinatorial
algorithm for the right mixed product
$C$ of $A$ with $B$
and its witness matrix 
is a generalization of that from \cite{BL}.
Let $P(r,v)$ denote a priority queue 
(implemented as a heap) on
the entries $A[r,k]$ such that $B[k,v]=1$
ordered by their values in nondecreasing
order. 

First, we compute
an approximate minimum spanning tree of the columns of $B$
in the Hamming metric. 
Then, we fix a traversal of the tree. 
Next, for each row $r$ of $A$, we traverse
the tree, construct $P(r,start)$
where $start$
is the first column of $B$
in the tree traversal
and then maintain
$P(r,v)$
for the currently traversed $v$ by
updating $P(r,u)$
where $u$ is the predecessor of $v$
in the traversal.
A minimum element in $P(r,v)$
yields a witness for $C[r,v]$.
The cost of the updates in a single
traversal of the tree is proportional
to the cost of the tree modulo
a logarithmic factor.

\vskip 0.5 cm
\noindent
\begin{center}
\Ovalbox{\small
\begin{minipage}[h]{0.99 \textwidth}
\centerline{\bf Algorithm 1}

{\bf Input:}
$n\times n$ matrix $A$ 
over $R\cup \{+\infty \}$ and an
$n\times n$ Boolean matrix $B;$
\\
{\bf Output:}
A witness matrix $W$ for
the right mixed product $C$ of $A$ and $B.$ 
\\
{\bf Comment}: $P(r,v)$ stands for a priority queue
on the entries $A[r,k]$ s.t. $B[k,v]=1$ ordered by their
values in nondecreasing order.
\\
\begin{enumerate}
\item Compute
an $O(\log n)$-approximate 
minimum spanning tree $T_B$ of 
the columns of $B$ in the Hamming metric;
\item Fix a traversal of the tree $T_B$
linear in its size;
\item Set $start$ to the first node
of the traversal;
\item For each
pair of consecutive neighboring
columns $v,\ u$ in the traversal, 
precompute the set $D_{v,u}$
of positions
where $1$s occur in $v$ but not
in $u$ and the set $D_{u,v}$
of positions where $1$s occur in
$u$ but not in $v;$
\item For each row $r$ of $A$ do
\begin{itemize}
\item Construct 
the priority queue
$P(r,start)$
and if $P(r,start)\neq \emptyset$ 
set $W[r,start]$ to the index $k$
where $A[r,k]$ is the minimum element in $P(r,start)$;  
\item Traverse the tree $T_B$
and for each node $v$ different
from $start$ compute 
the priority queue $P(r,v)$
from the priority queue $P(r,u)$,
where $u$ is the predecessor
of $v$ in the traversal, by
utilizing $D_{v,u}$ and $D_{u,v}.$
If $P(r,v)\neq \emptyset$
set $W[r,v]$ to the index $k$
where $A[r,k]$ is the minimum element in $P(r,v)$.
\end{itemize}
\end{enumerate}
\end{minipage}
}
\end{center}

\begin{lemma}\label{lem: 1}
Algorithm 1 is correct, i.e., it
outputs 
the witnesses matrix for
the right mixed product
of matrices $A$ and $B.$
\end{lemma}
For an $n\times n$ Boolean matrix $C,$
let $K_{C}$ be the complete weighted graph on 
the rows of $C$ where the weight
of an edge between two rows is equal to their Hamming distance.
Next, let $MWT(C)$ be the
weight of a minimum weight spanning tree of $K_{C}.$

\begin{lemma}\label{lem: 2}
Algorithm 1 
can be implemented in time
$\widetilde{O}(n(n + MWT(B^t))) +t(n),$  where
$t(n)$ is
the time taken
by the construction
of the $O(\log n)$-approximate
minimum weight spanning tree
in step 1.
\end{lemma}
\begin{proof}
Step 1 can be implemented in time
$t(n)$ while steps 2,3 take time $O(n).$
Step 4 takes $O(n^2)$ time. The block in Step 5
is iterated $n$ times. 

The first step in the
block, i.e., the construction of $P(r,start)$ takes
$O(n\log n)$ time. The update of $P(r,u)$
to $P(r,v)$ takes $O(\log n (|D_{v,u}|+|D_{u,v}|))$
time.
Note that $|D_{v,u}|+|D_{u,v}|$ is precisely
the Hamming distance between the columns $v$
and $u.$ It follows by the $O(\log n)$ approximation
factor of $T_B$ that the total time
taken by these updates is $O(MWT(B^t)\log^2 n).$

We conclude that Step 5 can be implemented 
in time $\widetilde{O}(nMWT(B^t)).$\qed
\end{proof}

\begin{theorem}\label{theo: mix}
The right mixed product of two $n\times n$ matrices
$A$ over $R\cup \{ +\infty\} $ and 
$B$ over $\{ 0,1\}$ can be computed 
by a combinatorial randomized algorithm
in time 
$\widetilde{O}(n(n + MWT(B^t))).$
Analogously, the left mixed product of $B$ and $A$
can be computed
by a combinatorial randomized algorithm
in time $\widetilde{O}(n(n + MWT(B))).$
\end{theorem}

\begin{proof}
By Corollary \ref{cor: Ham}, an $\Theta (\log n)$-approximate 
minimum spanning tree can be constructed
by a Monte Carlo algorithm
in time $\widetilde {O} (n^2)$ 
(observe  that $n^{1/f} = O(1)$ if $f = \Omega(\log n)$).
Hence, by Lemmata \ref{lem: 1}, \ref{lem: 2},
we obtain the theorem for the right
mixed product. The upper bound on the time
required to compute the left mixed product
follows symmetrically.\qed
\end{proof}

\section{Main results}

Lemma \ref{lem: mix} combined with Theorem \ref{theo: mix}
yield our main result.

\begin{theorem}\label{theo: main}
Let $G$ a directed graph $G$ on $n$ vertices
with nonnegative real vertex weights.
The all-pairs shortest path problem for $G$
can be solved by a combinatorial randomized
algorithm in time 
$\widetilde{O}(n^{2}\sqrt {n + \min\{MWT(A_G), MWT(A_G^t)\}}).$
\end{theorem}

By setting vertex weights, say, to zero, we obtain
immediately the following corollary.

\begin{corollary}\label{cor: tran}
The transitive closure of a directed graph $G$
on $n$ vertices can be computed by a combinatorial randomized
algorithm in time\\
$\widetilde{O}(n^{2}\sqrt {n + \min\{MWT(A_G), MWT(A_G^t)\}}).$ 
\end{corollary}

Equivalently, we can formulate Corollary \ref{cor: tran}
as follows.

\begin{corollary}\label{cor: tran2}
The transitive closure of an $n\times n$ Boolean
matrix $B$ (over the Boolean semi-ring)
can be computed by a combinatorial
randomized algorithm in time 
$\widetilde{O}(n^{2}\sqrt {n + \min\{MWT(B), MWT(B^t)\}}).$ 
\end{corollary}

\section{APSP in vertex-weighted uniform disk graphs of bounded density}

In this section, we consider
uniform disk graphs that are induced
by a set $P$ of $n$ points in a unit square in the plane that
are $b(n)$-dense, where $b: N\rightarrow N.$
Formally,
we say that $P$ is $b(n)$-dense  iff each cell of
the regular $\sqrt n \times \sqrt n$ grid within the unit square
contains at most $b(n)$ points.
The vertices of such an induced disk graph
are the points in $P$, and two vertices are adjacent in the graph
iff their Euclidean distance is at most $r,$ where $r$ is a 
positive constant not exceeding $1.$ 
We shall term
the aforementioned graphs as {\em uniform disk graphs 
induced by $b(n)$-dense point sets}.

\begin{lemma}
\label{lem:area}
Given two intersecting disks on the plane of the same radius $r$
with the distance $d$ between centers, the area of the symmetric
difference is $O(rd)$.
\end{lemma}
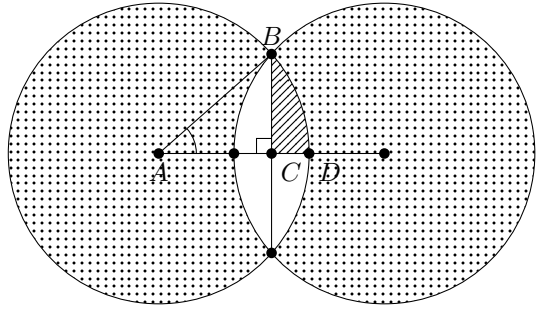
\begin{center}
\begin{tikzpicture}
\coordinate [label=below:$A$] (a) at (0,0);
\coordinate [label=above:$B$] (b) at (1.5, 1.322875656);
\coordinate [label=below right:$C$] (c) at (1.5,0);
\coordinate [label=below right:$D$] (d) at (2,0);
\coordinate (e) at (3,0);
\coordinate (f) at (1,0);
\coordinate (g) at (1.5, -1.322875656);
\draw (0.5,0) arc(0:41.41:0.5cm) (0.6,0.2) (a) circle(2) -- (e) circle(2);
\draw (a) -- (b) -- (c) -- (g) (1.5,0.2) -- (1.3,0.2) -- (1.3,0);
\fill (a) circle (2pt) (b) circle (2pt) (c) circle (2pt) (d) circle (2pt);
\fill (e) circle (2pt) (f) circle (2pt) (g) circle (2pt);
\begin{scope}
\path[clip] (d) arc(0:41.41:2) -- (c) -- cycle;
\path[pattern=north east lines] (a) circle (2);
\end{scope}
\begin{scope}
\path[clip] (b) arc(41.41:318.59:2) (g) arc(221.41:138.59:2);
\path[pattern=dots] (a) circle (2);
\end{scope}
\begin{scope}
\path[clip] (b) arc(138.59:-138.59:2) (g) arc(-41.41:41.41:2);
\path[pattern=dots] (e) circle (2);
\end{scope}
\end{tikzpicture}
\end{center}
\begin{proof}
$AC = \frac{d}{2}, AB = r$.

The area of the triangle $ABC$ is
$$Area_{ABC} = \frac{1}{2}AC BC
= \frac{1}{2} \frac{d}{2} \sqrt{r^2 - \frac{d^2}{4}} = \frac{1}{8} d \sqrt{4r^2-d^2}.$$

The area of the circular sector $ABD$ is 
$$Area_{ABD} =\frac{1}{2} r^2 \angle BAC
= \frac{1}{2}  r^2 \arccos \left(\frac{d}{2r}\right).$$

The area of $BCD$ is $Area_{BCD} = Area_{ABD} - Area_{ABC}.$

The area of the symmetric difference
$$Area= 2(\pi r^2 - 4 Area_{BCD}) =
2 \pi r^2 - 4 r^2 \arccos\left( \frac{d}{2r}\right) + d\sqrt{4r^2 - d^2} .$$

Finally, by using Taylor series expansion
\begin{multline*}
4r^2\arccos \left( \frac{d}{2r}\right) 
= 4r^2 \left(  \frac{\pi}{2} - \frac{d}{2r} + O\left(\left(\frac{d}{2r}\right)^2\right) \right)=
\\= 2\pi r^2 - 2dr + O(d^2) = 2\pi r^2 - 2dr + O(rd)
\end{multline*}
and  $\sqrt{4r^2-d^2} \leq 2r$ we get $ Area = O(rd).$
\qed
\end{proof}

\begin{lemma}\label{lem: eham}
Let $G$ be a 
uniform disk graph induced
by a $b(n)$-dense point set. For each edge $(v,u)$ of $G,$
the number of vertices in $G$ that are a neighbor of exactly one
of the vertices $v,\ u,$ i.e., the Hamming distance between
the two rows in the adjacency matrix of $G$ corresponding to $v$ and $u,$
respectively, is $O(r\times b(n)(dist(v,u)\times n + \sqrt n)).$
\end{lemma}
\begin{proof}
The number of vertices of $G$ that are a neighbor of exactly
one of the vertices $v$ and $u$ is at most the 
minimum number of cells of the regular $\sqrt n \times \sqrt n$
grid within the unit square that cover the symmetric difference
$S(v,u)$ between the disks centered at $v$ and $u,$ respectively,
multiplied by $b(n).$ The aforementioned number of cells is easily
seen to be at most the area $A(v,u)$ of $S(v,u)$ divided by the area
of the grid cell, i.e., $A(v,u)\times n$, plus the number of cells of
the grid intersected by the perimeter of $S(v,u)$, i.e., $O(r\sqrt n).$
By Lemma \ref{lem:area}, we have $A(v,u)=O(dist(v,u)\times r)$.
Hence, the aforementioned
number of cells is $O(r(dist(v,u)\times n + \sqrt n)).$

\qed
\end{proof}

The following lemma is a folklore (e.g., it follows directly
from the upper bound on the length of closed path through
a set of points in a $d$-dimensional cube
given in Lemma 2 in \cite{KS85}).

\begin{lemma}\label{lem: sqrt}
The minimum Euclidean spanning tree of any set of $n$ points
in a unit square
in the plane has total length $O(\sqrt n).$
\end{lemma}

Combining Lemmata \ref{lem: eham}, \ref{lem: sqrt}, we obtain the following one.

\begin{lemma}\label{lem: rug}
For a uniform disk graph $G$ 
induced by a $b(n)$-dense $n$-point set,
a spanning tree of
the rows (or, columns) of the adjacency matrix of $G$
in the Hamming metric  having cost $O(rn^{3/2})$
can be found in time $O(n^2).$
\end{lemma}

\begin{proof}
Construct a minimum Euclidean spanning tree of
the $n$ points forming the vertex set of $G.$
It takes time $O(n\log n)$ and the resulting
tree $T$ has total length $O(\sqrt n)$ by Lemma \ref{lem: sqrt}.
Form a spanning tree $U$ of the rows (or, columns) of
the adjacency matrix of $G$ by connecting by edge the
rows corresponding to $v$ and $u$ iff $(v,u)\in T.$
By Lemma \ref{lem: eham} and the $O(\sqrt n)$ length of $T,$
the total cost of $U$ is $O(rn^{3/2}b(n)).$
\qed
\end{proof}

By plugging Lemma \ref{lem: rug} into Theorem \ref{theo: main}, we obtain
our main result in this section.

\begin{theorem}\label{theo: amain}
Let $G$ be a uniform disk graph, with nonnegative
real vertex weights, 
induced by a $b(n)$-dense $n$-point set.
The all-pairs shortest path problem for $G$
can be solved by a combinatorial
algorithm in time 
$\widetilde{O}(\sqrt r n^{2.75}\sqrt {b(n)}).$
\end{theorem}

In the application of the method of Theorem \ref{theo: main}
yielding Theorem \ref{theo: amain}, we can use the
deterministic algorithm of Lemma \ref{lem: rug} 
to find a spanning tree of the rows or columns
of the adjacency matrix of $G$ instead of the
randomized approximation algorithm from Fact 2.

By straightforward calculations,
our upper time-bound for APSP in
vertex-weighted uniform disk graphs 
induced by $O(1)$-dense point sets
subsumes that for APSP in sparse graphs
based on Dijkstra's single-source shortest-path
algorithm, running in time $\widetilde{O} (nm),$
where $m$ is the number of edges, for $r>>n^{-1/6}.$

\junk{We can immediately extend Theorem \ref{theo: amain} to include
the directed case just by requiring the underlying
undirected graph of $G$ to be a 
uniform disk graph of $b(n)$-density.} 

Finally, we can also easily extend 
Theorem \ref{theo: amain} to include uniform ball
graphs in a $d$-dimensional Euclidean space. 
In the extension, the term $\sqrt r$ in the upper
time-bound generalizes to $\sqrt {r^{d-1}}.$

\section{Final Remarks}

We can easily extend our main result to include
solving the APSP problem for vertex and edge weighted
directed graphs in which the number of different edge weights
is bounded, say by $q.$ This can be 
simply achieved by decomposing the adjacency matrix
$A_G$ into the union of up to $q$ matrices $A_1,A_2,...A_l$
in one-to-one
correspondence with the distinct edge weights and consequently
replacing each mixed product with $l$ such products
in Lemmata \ref{lem:i+1}, \ref{lem: mix}.
In the final upper bound, $MWT(A_G)$ and $MWT(A_G^t)$
are replaced by $\sum_{i=1}^l MWT(A_i)$ and $\sum_{i=1}^l MWT(A_i^t),$
respectively.

It is an interesting problem to determine if there are
other natural graph classes where $MWT(A_G)$ or $MWT(A_G^t)$
are substantially subquadratic in the number of vertices.

It follows from the existence of the so called Hadamard matrices
\cite{C} that there is an infinite sequence of graphs with
$n_i\times n_i$ adjacency
matrices $A_i$ such that 
\junk{there is a set of $\Omega(n_i)$ rows of $B_i$
where the Hamming distance between any pair of rows is $\Omega(n_i)$
as well as a set of $\Omega(n_i)$ columns of $B_i$ where the Hamming
distance between any pair of columns is $\Omega(n_i)$. Note that for
such matrices $B_i$, which can be interpreted as adjacency matrices of
the corresponding directed graphs,}
 $\min\{MWT(A_i), MWT(A_i^t)\}=
\Omega ((n_i)^2)$ holds.

\junk
{In \cite{GL}, the
upper time-bound on Boolean matrix product
from \cite{BL} is strengthened by showing that one can
replace the Hamming distance
with the so called extended Hamming distance
in the definition of the cost of the minimum
spanning tree of matrix rows or columns.
The extended Hamming distance between two $0-1$ strings
never exceeds the Hamming one as it counts
the differences between corresponding blocks of
consecutive zeros and ones. Unfortunately,
an analogous strengthening of our upper time-bound
on mixed matrix product does not seem possible.

On the other hand, it seems that whenever the rows of 
a Boolean matrix $B$ or its columns
admit a substantially subquadratic representation, there
might be good chances for computing 
one of the mixed product of $B$ with a matrix over $R\cup \{+\infty\}$
combinatorially
in substantially subcubic time. 

On the other hand, the
absence of such representations might mean that the matrix $B$
has some properties of a random one and therefore could admit
a substantially subcubic combinatorial 
algorithm for their Boolean product
like the random ones \cite{SS1}. This general idea gives
some hope in the search for a combinatorial substantially
subcubic algorithm for Boolean matrix product.

\section{Acknowledgments}

The authors are grateful to unknown referees
for valuable comments, and to
the whole algorithm group at Kiel university
and Miroslaw Kowaluk
for discussions on related shortest path problems.}

\end{document}